\documentclass[11pt,a4paper]{article}
\usepackage{amsmath,amssymb}
\usepackage{natbib}
\usepackage{graphicx}
\usepackage{lineno}
\usepackage{hyperref}
\usepackage{setspace}
\usepackage[hmargin=3cm]{geometry}
\title{Semi-brittle flow of rocks: Cracks, dislocations and strain hardening}
\author{Nicolas Brantut\\Department of Earth Sciences\\University College London, London, UK}
\date{\ }

\newenvironment{eq}{\begin{linenomath}\equation}{\endequation\end{linenomath}}
\begin{document}

\maketitle

\begin{abstract}
  Strain hardening is a key feature observed in many rocks deformed in the so-called ``semi-brittle'' regime, where both crystal plastic and brittle deformation mechanisms operate. Dislocation storage has long been recognised as a major process leading to strain hardening. Here, we suggest that tensile microcracks may be viewed as dislocation sinks, by offering internal free surfaces where dislocations can escape individual crystals within an aggregate. Strain hardening is modelled with a conventional approach, combining Taylor's equation relating stress to dislocation density, and a dislocation density evolution law based on dislocation mean-free path and dynamic recovery. The initiation of microcracks is modelled as a function dislocation density, assuming dislocation pile-ups at grain boundaries. Microcrack growth is modelled using linear elastic fracture mechanics. The model captures important qualitative features observed in calcite marble deformation experiments: pressure-dependency of strength in the ductile regime, and a reduction in hardening linked to an increase in crack growth with decreasing confining pressure. Grain-size dependency of strength and hardening is also captured but requires significant toughening (or limitation to crack growth) at small grain sizes. The model can be improved significantly once detailed, systematic microstructural observations become available.
\end{abstract}

\section{Introduction}

With increasing pressure and temperature, rocks transition from a brittle behaviour marked by microcracking that leads to macroscopic failure, to a fully plastic behaviour where deformation can be accomodated in a distributed fashion by crystal plastic processes such as dislocation motion and solid-state diffusion \citep[e.g.][]{evans90}. In the transitional regime, at pressures sufficiently high to suppress macroscopic faulting, and temperatures sufficiently low to limit diffusion, most rocks behave in a semi-brittle, ductile fashion \citep[e.g.][]{evans90,evans95,kohlstedt95}. Deformation is accomodated by a combination of microscopic slip at interfaces (shear cracks), tensile microcracks, and crystal plastic processes such as twinning and dislocation motion. This behaviour is well documented in rocks like calcite marble \citep{fredrich89,rybacki21}, quartzite \citep{hirth94} and peridotite \citep{druiventak11}.

Strain hardening is a key feature of ductile rock deformation in the semi-brittle regime: under typical laboratory conditions and sample sizes, flow stress does not reach a clear steady state \citep[e.g.][]{rybacki21}, which implies that microstructures keep evolving. Strain hardening is now recognised to be an important phenomenon that controls the strength of the lithosphere \citep[e.g.][]{hansen19,meyer19,wallis20}, but its physical underpinning is not fully understood.

In general, strain hardening is produced by microstructural changes that either increase the internal elastic stress in the material, or raise the critical stress required to drive the motion of carriers of deformation \citep[e.g.][]{mecking81,kocks03}. In the semi-brittle regime of rocks, two mains hardening mechanisms can be distinguished: (1) increase in elastic restoring forces (sometimes also called backstresses) due to slip across preexisting defects (e.g., cracks) in an otherwise elastic matrix \citep[e.g.][]{lawn98,david20}, (2) increase in strength due to accumulation of stored dislocations in deforming crystals \citep[e.g.][]{taylor34}. Slip across preexisting defects (in an otherwise elastic matrix) is a general process and is not restricted to dislocation glide: frictional slip is also a potentially important phenomenon at low temperature and low pressure in many rock types \citep[e.g.][]{david20,brantut23,liu23}. Twinning can also produce backstress \citep[e.g.][]{mitchell91}.

%The effect of dislocation glide and storage on strain hardening has been well studied in the context of metals, and is well captured by 

In the semi-brittle regime, the role of tensile microcracks on strength is not clear. Microcracks do not strongly interact to produce a localised macroscopic shear fault, but they produce significant dilation \citep[e.g.][]{edmond72,fredrich89}. In addition, there is an inverse dependency between the degree of microcracking and the hardening rate; experiments in calcite marble have shown that higher crack densities are associated with lower hardening rates \citep{fredrich89,harbord23}. Only a few micromechanical models have been developed where tensile crack growth is coupled to plastic flow. The early approach of \citet{horii86} was based on an extension of the ``wing crack'' model (originally developed to understand crack growth in brittle solids), and included shear flaw extension into plastic zones. Recent reexamination of this model \citep{liu23} showed that this approach leads to useful insights into the onset of microcracking in the semi-brittle regime, but does not appropriately reproduce the full stress-strain behaviour. Similary, the model of \citet{renshaw01} introduces plastic relaxation near crack tips in an otherwise brittle model, which is useful to make predictions of fracture strength, but not to stress-strain behaviour. By contrast, \citet{nicolas17} presented a coupled model, including crack growth and dislocation glide, aimed to predict stress-strain behaviour, dilatancy, compaction and crack density evolution during high pressure deformation of porous limestone. While quite succcessful, the model of \citet{nicolas17} is made very complex due to the need to account for frictional processes and void enlargement/compaction in addition to plastic flow and tensile cracking.

%somewhat challenging to use. Nevertheless, one key ingredient

The goal of this note is to develop a simple model to explain how microcracking may be coupled to dislocation glide to produce some of the weakening effect observed in experiments. Our starting point will be the ``one state variable'' flow law of \citet{mecking81,estrin84}, where dislocation density determines the full evolution of stress-strain behaviour. This model, which we will refer to as the ``Kocks-Mecking-Estrin'' (abbreviated KME) model, has proved very successful to understand the behaviour of many metals \citep[see review by][]{kocks03}. It has also been extended to include the role of twins in metals \citep[e.g.][]{bouaziz08,decooman18}, which may also apply to rock-forming minerals \citep{rybacki21,harbord23}. Here, we propose to include the effect of microcracks on dislocation density evolution by considering that open cracks acts as dislocation sinks, so that an increased crack density leads to a decrease in strain hardening, as observed experimentally. The nucleation and growth of tensile cracks will be coupled to dislocation density evolution by considering that cracks nucleate from dislocation pile-ups, an approach that was used by \citet{nicolas17}. While much experimental work is still needed to confirm the validity of the approach, the present work provides a first quantitative attempt to explain some key aspects of semi-brittle flow in rocks, and testable predictions are suggested.

\section{Model}

\subsection{Dislocation-based strain hardening process}

The KME model is based on the Taylor relationship \citep{taylor34} relating the macroscopic shear stress $\tau$ to dislocation density $\rho$ as
\begin{eq} \label{eq:taylor}
  \tau = \tau_0 + M \alpha b G \sqrt{\rho},
\end{eq}
where $\tau_0$ is a constant minimal stress required to move dislocation (lattice friction), $\alpha$ is a constant of $O(1)$, $M$ is a geometrical factor related to activation of multiple slip systems in different orientations (sometimes called ``Taylor'' factor), $b$ is the magnitude of the Burgers vector representative of the characteristic slip system producing hardening and $G$ is the shear modulus. The term $M \alpha b G \sqrt{\rho}$ corresponds to a ``backstress'', i.e., the additional stress required to move dislocations, which arises from dislocation interactions (e.g., dislocation tangles). The relationship \eqref{eq:taylor} has been shown to be reasonably accurate in most rock forming minerals deformed in the low temperature regime, including quartz \citep{kohlstedt80}, and calcite \citep{debresser96}, plagioclase and olivine \citep{thom22}. It is also applicable in polycrystalline aggregates, at least at relatively high stress \citep{debresser96}. In olivine, recent work by \citet{breithaupt23} has shown that the Taylor relationship \eqref{eq:taylor} is applicable provided that the $\tau_0$ term is properly accounted for, which was not necessarly recognised in previous studies \citep{kohlstedt74,durham77,bai92}.

The Taylor equation \eqref{eq:taylor} includes the average effect of dislocations on the stress required to deform the material. There might be other contributions in the form of backstress, i.e., elastic restoring forces appearing due to distorsion of crystals during deformation, that can be accounted for by modifying Equation \eqref{eq:taylor} as follows \citep[e.g.][]{sinclair06}:
\begin{eq} \label{eq:taylor_b}
  \tau = \tau_0 + M \alpha b G \sqrt{\rho} + \tau_\mathrm{b},
\end{eq} 
where $\tau_\mathrm{b}$ denotes the additional backstress that could arise from other mechanisms or specific dislocation configurations. Experimental evidence for the existence of backstress is in the form of the Bauschinger effect, where material yields at a different threshold during reversal of deformation direction: the backstress ``helps'' yielding during strain reversal because it acts as a reaction force that opposes the initial loading direction\footnote{This effect is the analogue of the stress-memory effect due to frictional slip in brittle rocks \citep[e.g.][]{holcomb81,brantut23}.}. Such an effect has been observed in olivine \citep{hansen19,hansen21,thom22} and may be relevant in other rock forming minerals.

Here we follow \citet{sinclair06} and consider that backstress arises from dislocation pile-ups at obstacles such as grain boundaries. For an isolated dislocation pile-up of length $L$ containing $n_\mathrm{p}$ dislocations of Burgers vector $b$, elastic equilibrium requires that the driving stress (here equivalent to our backstress) is \citep{eshelby51}
\begin{eq} \label{eq:taub}
  \tau_\mathrm{b} = \frac{G}{\pi(1-\nu)}\frac{n_\mathrm{p}b}{L},
\end{eq}
where $\nu$ is Poisson's ratio. In general, Equation \eqref{eq:taub} should represent an upper bound for the net, average backstress in a polycrystalline aggregate: while the expression applies locally to an isolated dislocation pile-up, the net effect on the average stress at aggregate level may be much smaller due to interactions and superposition of pile-ups of different signs and orientations.

The number of dislocations in pile-ups, $n_\mathrm{p}$, is expected to be connected to the total number of dislocations in the material, represented by the dislocation density $\rho$. In a hardening model including backstress from dislocation pile-ups, \citet{sinclair06} suggested an evolution law for $n_\mathrm{p}$ such that $n_\mathrm{p}$ saturates at some characteristic upper bound, which is meant to represent the relaxation of elastic stress concentrations due to stabilisation of pile-ups by dislocations of opposite signs.

Here, in absence of detailed observations in rock-forming minerals, an alternative, simpler approach is proposed. Any additional backstress $\tau_\mathrm{b}$ can only exist if (1) there are sufficiently ``opaque'' barriers to dislocation motion, and (2) there is an imbalance in net dislocation signs across barriers. It is possible to accumulate (store) dislocations without generating significant backstress, as long as the stored dislocations are in a stable configuration: therefore, only a fraction of the total dislocation density is expected to be present in ``unbalanced'' pile-ups. If we denote this fraction $\phi$, we can then write the number of dislocations in each pile-up as
\begin{eq} \label{eq:nb}
  n_\mathrm{p} = \phi \rho / \rho_\mathrm{pu},
\end{eq}
where $\rho_\mathrm{pu}$ is the pile-up density (defined as in \citet{nicolas17}, $\rho_\mathrm{pu} = 1/L_\mathrm{pu}$ where $L_\mathrm{pu}$ is the spacing between pile-ups).

The total dislocation density $\rho$ is the internal state variable that determines the stress evolution during deformation. Based on extensive experimental observations in metals deformed at low temperature, the evolution of $\rho$ with plastic strain has been found to be well captured by a relation of the form \citep[e.g.][]{kocks03}
\begin{eq} \label{eq:drhode}
  \frac{d\rho}{d\epsilon} = \frac{1}{b\lambda} - f\rho - R(T, \dot{\epsilon}, \ldots),
\end{eq}
where $\epsilon$ is the shear strain, $\lambda$ is the dislocation mean free path, $f$ is a dynamic recovery parameter (that depends weakly on temperature and strain rate), and $R$ is a collective term denoting other potential recovery processes \citep[e.g.][]{breithaupt21,breithaupt23}. Equation \eqref{eq:drhode} is written in terms of strain increments, following the requirement that dislocation nucleation is due to imposed deformation. A more general approach, detailed in \citet{breithaupt21}, would involve a evolution law for $\rho$ with increments of time, which is a more natural independent variable when considering static recovery processes and rate-dependent dislocation creep. Here, we use the form \eqref{eq:drhode} because we focus on the strain-hardening behaviour of rocks at low temperatures, where rate effects are minor \citep[e.g.][]{rybacki21}.

The dislocation mean free path is impacted by two major obstacles: grain boundaries, and other dislocations. A common form for $\lambda$ is
\begin{eq} \label{eq:lambda}
  1/\lambda = 1/d + k\sqrt{\rho},
\end{eq}
where $d$ is the grain size and $k$ is a constant. The term $k\sqrt{\rho}$ describes so-called ``forest hardening'', i.e., the potential for existing dislocations to ``pin'' the motion of other dislocations and produce tangles.

The combination of Equations \eqref{eq:taylor}, \eqref{eq:drhode} and \eqref{eq:lambda} has been very effective in understanding the strain hardening behaviour in many materials where dislocations are the main carrier of deformation \citep{kocks03}. What is proposed here is an extension of this approach where the existence of cracks is accounted for.

\subsection{Crack nucleation and growth}

In crystals where dislocation glide is active but not on sufficiently many slip systems to produce fully plastic flow, dislocations may pile up at obstacles and raise internal stress, which leads to microcracking. Cracking associated to dislocation glide has been studied by \citet[][among others]{stroh54,stroh58}. The effect of pressure on crack nucleation during plastic flow was established by \citet{francois68} and studied more systematically by \citet{wong90}. The model of \citet{wong90} was used recently in the semi-brittle micromechanical model of \citet{nicolas17}.

As a note of caution, it should be stated that direct experimental support for microcracking due to dislocation pile-up remains at best incomplete in geological materials. Potential observations have been discussed in the original work of \citet{wong90}. Cracks are often associated with dislocations in rocks deformed in the semi-brittle regime \citep{darot85,fredrich89,druiventak11,wallis17}, but it is not clear whether cracking is induced by dislocations. Despite the uncertainty of the fracturing mechanism, cracks are observed to interact with dislocations in many instances \citep[e.g.][]{fredrich89}, and in absence of solid alternative processes we will follow \citet{nicolas17} and work on the hypothesis that dislocation pile-ups are one of the main potential source of microcracks in the semi-brittle regime.

The condition for crack nucleation ahead of a pile-up is given by \citep{wong90}
\begin{eq}
  K_\mathrm{Ic} = \frac{4\sqrt{2}}{\sqrt{3\pi}}\tau_\mathrm{eff}\sqrt{L},
\end{eq}
where $K_\mathrm{Ic}$ is the fracture toughness, $L$ is the length of the pile-up (that should be commensurate to the grain size, or, to the dislocation mean free path), and $\tau_\mathrm{eff}$ is the stress driving the dislocations in the pile-up. The number of dislocations in the pile-up is given by
\begin{eq}
  n_\mathrm{p} = \frac{\pi(1-\nu)L\tau_\mathrm{eff}}{b G},
\end{eq}
so the nucleation condition is satisfied if \citep{nicolas17}
\begin{eq}
  n_\mathrm{p}>n_\mathrm{c} = \frac{\pi\sqrt{3\pi L}}{8bG/(1-\nu)}K_\mathrm{Ic}.
\end{eq}

\begin{figure}
  \centering
  \includegraphics{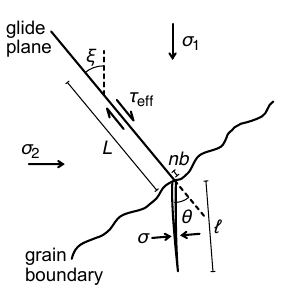}
  \caption{Schematic of a crack nucleated at a pile-up. Modified from \citet{olsson76,wong90}.}
  \label{fig:pileupcrack}
\end{figure}

Once a crack is nucleated at a pile-up, it will grow to maintain the condition $K_\mathrm{I}\leq K_\mathrm{Ic}$. The mode I stress intensify factor at the tip of the open crack is given by \citep{wong90}
\begin{eq} \label{eq:crackgrowth}
  K_\mathrm{I} = n_\mathrm{p}b\frac{G}{1-\nu}\frac{\sin\theta}{\sqrt{2\pi\ell}} - \sigma\sqrt{\pi\ell/2},
\end{eq}
where $\theta$ is the angle between the pile-up plane and the open crack, $\ell$ is the open crack length, and $\sigma$ is the normal stress applied on the open crack (positive in compression), expressed as (see notations and geometry in Figure \ref{fig:pileupcrack})
\begin{eq}
  \sigma = \frac{\sigma_1+\sigma_2}{2} - \frac{\sigma_1-\sigma_2}{2}\cos\big(2(\theta-\xi)\big).
\end{eq}

%In any case, as stated by \citet{wong90}, the details of the model geometry do not seem to be of primary importance

\subsection{Feedback between cracking and dislocation storage}

If $n_\mathrm{p}>n_\mathrm{c}$, cracks will nucleate at pile-ups. How would this impact strength? One may think that cracking would relax the backstress. This is true locally at the head of the pile-up, but tensile cracking does not suppress elastic stresses: it merely displaces the stress concentration from the original pile-up front to the tensile crack front. There is always an elastic restoring force associated with the slip accumulated along the pile-up plane \citep[see analogous case of the wing crack, e.g.][]{basista98}. The backstress could be reduced if the tensile crack tip is shielded by plastic deformation, e.g., by a cloud of dislocations \citep[][section 7.3]{lawn93}.

The open microcracks provide free surfaces inside the material. Free surfaces act as dislocation sinks \citep[][Chap. 9]{caillard03}: mobile dislocations reaching free surfaces generate steps, and the dislocation line disappears from the crystal. The dislocation density change due to the presence of free surfaces can be written \citep[][Equation 9.6]{caillard03}
\begin{eq} \label{eq:drho_free}
  \left(\frac{d\rho}{d\epsilon}\right)_\mathrm{free} = -\frac{1}{b\lambda_\mathrm{c}},
\end{eq}
where $\lambda_\mathrm{c}$ is the characteristic spacing between cracks along the dislocation glide plane. Considering an array of parallel cracks of length $\ell$ spaced on average by the distance $L_\mathrm{c} = 1/\sqrt{N_\mathrm{A}}$, we can estimate the mean distance for a dislocation gliding perpendicular to the cracks to reach a crack plane by
\begin{eq}
  \lambda_\mathrm{c} = L_\mathrm{c}^3/\ell^2.
\end{eq}
We can verify that $\lambda_\mathrm{c}=\infty$ for $\ell=0$ (i.e., dislocations never reach a crack surface if there are no cracks), and $\lambda_\mathrm{c}=L_\mathrm{c}$ for $\ell=L_\mathrm{c}$ (i.e., dislocation travel an average distance $L_\mathrm{c}$ to reach a crack surface if the crack size is equal to the crack spacing).

Including the contribution \eqref{eq:drho_free} in the dislocation evolution model, we obtain
\begin{eq}
  \frac{d\rho}{d\epsilon} = \frac{1}{b}\left(\frac{1}{d} - \frac{\ell^2}{L_\mathrm{c}^3} + k\sqrt{\rho}\right) - f\rho - R(T,\dot\epsilon,\ldots),
\end{eq}
where we see how crack growth might lead to a reduction in dislocation storage rate and therefore in hardening rate.

\subsection{A remark on twinning}

The model as derived above only includes the effect of dislocations and microcracks on strength and hardening. We would like to test the model predictions and compare them to experimental data on rocks. The rock type that has been most extensively studied in the semi-brittle regime is calcite marble (more specifically, Carrara marble): this material is known to easily twin during deformation \citep[e.g.][among many others]{barber79,spiers79,rowe90}, and twinning has been suggested to contribute significantly to strain hardening \citep{rybacki21}. In this section, we discuss an extension of the model that includes the role of twins. This is however not an essential element of the analysis which aims to focus on the role of cracks, and this is only provided for completeness and an improved quantitative analysis of data obtained on calcite.

The KME model has been successfully extended to account for the presence of twins \citep[e.g.][]{bouaziz08}. One key feature of the stress-strain behaviour of metals that produce twins during deformation is the large hardening rates, orders of magnitude larger than typical metals \citep{decooman18}. This behaviour may be explained by the fact that twin boundaries are obstacles to dislocation motion, and thus considerably reduce the dislocation mean free path $\lambda$. This effect can be accounted for by expressing \citep[e.g.][]{bouaziz08}
\begin{eq} \label{eq:lambda_twin}
  1/\lambda =  1/d + 1/t + k\sqrt{\rho},
\end{eq}
where $t$ is the twin spacing. While an expression like \eqref{eq:lambda_twin} has been very successful in explaining the behaviour of twinning-induced plasticity in steel, its validity has not yet been thoroughly tested in rocks such as calcite (which has an easy twin system active at low temperature). Recent experimental data on calcite marble by \citet{rybacki21} suggest that the approach remains valid, but  \citet{harbord23} have obtained results that indicate that the effect of grain size might be dominant over the effect of twin density in controlling hardening rate. The relatively minor role of twinning compared to that of grain size might be explained by the ease of slip transmission across twin boundaries in calcite \citep{harbord23}. In general, we expect different types of obstacles to contribute differently to the dislocation mean free path, depending on their effectiveness in stopping dislocations. We could thus write
\begin{eq} \label{eq:lambda_weights}
  1/\lambda =  k_\mathrm{d}/d + k_\mathrm{t}/t + k\sqrt{\rho},
\end{eq}
where $k_\mathrm{d}$ and $k_\mathrm{t}$ are weighting factors that reflect the ``opacity'' of each barrier types. In practice, there is currently insufficient experimental constraints to warrant a detailed breakdown of such empirical constants, and grain size and twin density effects can be lumped into a single ``effective'' quantity $d_\mathrm{eff}$ such that
\begin{eq}
  1/d_\mathrm{eff} = k_\mathrm{d}/d + k_\mathrm{t}/t.
\end{eq}
Furthermore, in the case of calcite, $d_\mathrm{eff}$ can be approximated as a constant, since twin density is only weakly dependent on strain in this material \citep{rowe90,rybacki13}.

\section{Parameters and unknowns}

The model contains a number of parameters that need to be constrained experimentally. This is a challenging task because only a few datasets exist that relate stress, strain and dislocation density together with semi-brittle flow in rocks. The prime candidate is calcite marble, where the most complete dataset exist in the regime of interest.

\begin{figure}
  \centering
  \includegraphics{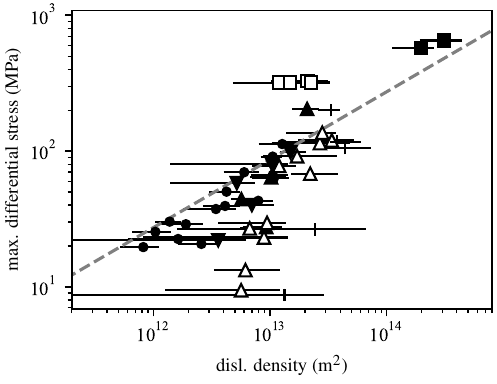}
  \caption{Stress vs. dislocation density in calcite, as complied and corrected originally by \citet{debresser96}. Dashed line is Equation \eqref{eq:taylor} with constant modulus $G=32$~GPa, $b=6.37$~\AA and $M\alpha=1.34$. Symbols are : $\bullet$ single crystals deformed at $T\geq550^\circ$C \citep{debresser96}; $\square$ Carrara marble deformed at room temperature and $P_\mathrm{c}$ ranging from 120 to 300~MPa \citep{fredrich89}; $\blacktriangle$ Carrara marble deformed at $P_\mathrm{c}=300$~MPa and temperatures from 600 to 1000$^\circ$C \citep{debresser96}; $\blacktriangledown$ Yule marble deformed at 500~MPa and 600 to 800$^\circ$C \citep{goetze77}; $\blacksquare$ Solnhofen limestone deformed at room temperature and $P_\mathrm{c}=200$ to $220$~MPa \citep{briegel78}; $\vartriangle$ Carrara marble deformed at 300~MPa and $600$ to $1050^\circ$C \citep{schmid80}; $+$ Solnhofen limestone deformed at 300~MPa and $600$ to $900^\circ$C \citep{schmid77}.}
  \label{fig:debresser_fredrich_disl}
\end{figure}

For calcite single crystals as well as polycrystals at high stress and high temperature ($>550^\circ$C), the Taylor relationship \eqref{eq:taylor} adequately fits in the high stress regime (Figure \ref{fig:debresser_fredrich_disl}), with parameters $M\alpha=1.34$, $b=6.37$~\AA, and an average shear modulus $G=32$~GPa \citep{debresser96}. The good fit with most polycrystalline aggregates also implies that specific backstress due to pile-ups play a minimal role, and could potentially be neglected. Note that dislocation densities measured in samples deformed at room temperature \citep{fredrich89} do not strictly follow the trend of single crystals, with stress higher than expected. It is possible that the additional stress at a given dislocation density arises from backstress due to dislocation pile ups or twins, or from frictional processes that could still be significant at the low pressure (around $100$~MPa) used in \citet{fredrich89} experiments. A discussion of the behaviour at low stress, not directly relevant to our model of semi-brittle flow, can be found in \citet{debresser96}.

\begin{figure}
  \centering
  \includegraphics{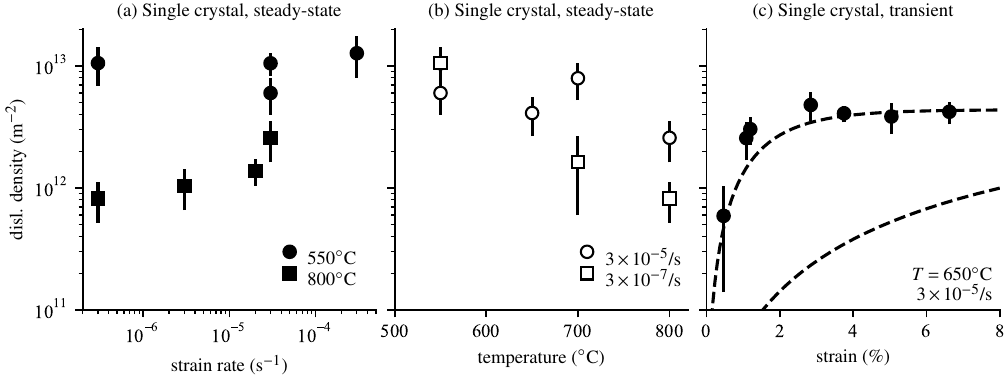}
  \caption{Dislocation density in calcite single crystals as measured by \citet{debresser96}. Plots (a) and (b) are steady-state data at different temperatures and strain rates, and plot (c) show variations in dislocation density as a function of strain. Dashed line in (c) is a fit to Equation \eqref{eq:rho_single} with $k=0.2$, $f=150$ and $\rho_0=10^{10}$~m$^{-2}$.}
  \label{fig:debresser_singlecrystal}
\end{figure}

If we consider single crystals, thus neglecting grain size effects, it is possible to estimate parameters $k$ and $f$ from the data compiled in \citet{debresser96} (Figure \ref{fig:debresser_singlecrystal}).  Neglecting size effects ($d\gg1$) and assuming no additional static recovery effects, the solution of \eqref{eq:drhode} is
\begin{eq} \label{eq:rho_single}
  \rho(\epsilon) = \left[e^{-f\epsilon/2}\left(\sqrt{\rho_0}-k/(bf)\right) + k/(bf) \right]^2,
\end{eq}
where $\rho_0$ is the initial dislocation density. The steady-state dislocation density is $\rho_\mathrm{ss} = \big(k/(bf)\big)^2$. Both steady-state and transient dislocation density measurements (Figure \ref{fig:debresser_singlecrystal}) can be fitted by Equation \eqref{eq:rho_single}; we obtain $k=0.2$, and $f$ of the order of $100$ at low temperature ($550^\circ$C). Both these parameter values seem quite high compared to those typically obtained in metals, where $k\sim0.01$ and $f\sim1$ \citep[e.g.][]{bouaziz01}. One possible reason for the apparent large values of $k$ and $f$ is the existence of additional recovery processes not included in Equation \eqref{eq:rho_single}, notably dislocation climb. Therefore, the estimates based on single crystal data at elevated temperature (Figure \ref{fig:debresser_singlecrystal}) should be viewed with caution, and we will investigate the behaviour of the system with a wider range of parameter values.

The full model including microcracks requires the use of several additional parameters. The fracture toughness is quite well constrained in many rock forming minerals \citep{atkinson84}, and is $K_\mathrm{Ic}=0.2$~MPa~m$^{1/2}$ for calcite at room temperature \citep{santhanam68,atkinson80b}. By constrast, the quantities like pile-up density $\rho_\mathrm{pu}$, pile-up length $L$, microcrack spacing $L_\mathrm{c}$, and the fraction of dislocations stored in pile-ups $\phi$ are yet to be constrained by any experimental data or observations. Due to the simplified nature of the model, it is unlikely that direct measurements can be made to identify these parameters with certainty. Here, we have to rely on a general assessment in terms of order of magnitude, and make elementary predictions. Following \citet{nicolas17}, we will use $\rho_\mathrm{pu}=10^8$~m$^{-2}$, $L=d$ (grain size) and $\phi$ up to $1$. The nominal crack spacing $L_\mathrm{c}$ can also be taken equal to the grain size, since we assume that crack nucleate at pile-ups. One final parameter required by the model is related to the level of backstress: as noted earlier, the expression \eqref{eq:taub} is the backstress due to an isolated dislocation pile-up, and the average backstress in the aggregate is most likely much smaller than this value. To reflect this fact, we introduce an ad-hoc factor $f_\mathrm{b}\ll1$ that multiplies $\tau_\mathrm{b}$, and set it to a constant. Table \ref{tab:param} summarises the parameter values used to model calcite deformation.

\begin{table}
  \centering
  \caption{List of model parameters relevant to calcite aggregates.}
  \label{tab:param}
  \begin{tabular}{lc}
    \hline
    Parameter & Value\\
    \hline
    Taylor factor, $\alpha M$ & 1.34\\
    Burgers vector, $b$ & 6.37 \AA\\
    Shear modulus, $G$ & 32 GPa\\
    Poisson's ratio, $\nu$ & 0.28\\
    Forest hardening parameter, $k$ & $0.2$\\
    Dynamic recovery parameter, $f$ & $0$--$200$\\
    Backstress factor, $f_\mathrm{b}$ & $0.01$ \\ 
    Effective grain size, $d_\mathrm{eff}$ & 150 $\mu$m\\
    Fraction of dislocations in pile-ups, $\phi$ & 0.1--1\\
    Pile-up density, $\rho_\mathrm{pu}$ & $10^8$~m$^{-2}$\\
    Fracture toughness, $K_\mathrm{Ic}$ & 0.2~MPa/m$^{1/2}$\\
    Crack angle, $\theta$ & $\pi/4$\\
    \hline
  \end{tabular}
\end{table}

\section{Results}
\subsection{General features}

\begin{figure}
  \centering
  \includegraphics{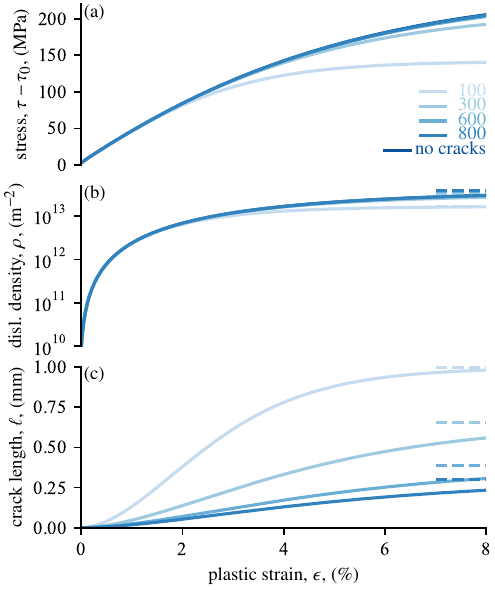}
  \caption{Model prediction for stress variation, dislocation density and tensile crack length with increasing plastic strain for a range of confining pressures (values given in MPa in panel (a)). Parameter values are given in Table \ref{tab:param}, and $f=50$. Dashed lines in panels (b) and (c) indicate steady-state solutions.}
  \label{fig:model_example}
\end{figure}

The model predicts a nonlinear strain hardening behaviour that depends on confining pressure (taken as $\sigma_2$ in the model; Figure \ref{fig:model_example}). At low pressure, cracks can grow easily and efficiently limit dislocation storage. With increasing pressure, crack growth is limited and more dislocations are stored, leading to significant hardening. A steady-state in terms of dislocation density, crack length and stress can be achieved when the sink terms (due to cracking and dynamic recovery) match the source terms (due to grain size and forest hardening). With the large parameter $f$ inferred from single crystal data, of the order of $100$, the steady state is achieved after around 5\% strain.

When cracking is not accounted for, there is no predicted pressure dependency: strain hardening continues until the steady-state $\rho=(k/(bf))^2$ is reached. In practice, at low temperature, this steady-state may correspond to very large dislocation densities, and it is not clear if this situation is realistic in absence of dislocation climb processes.

The model output is strongly influenced by all the newly introduced parameters: with increasing $\phi$, microcrack growth initiates at lower strain and produces an earlier stabilisation of the dislocation density. With increasing $f_\mathrm{b}$, the strain hardening rate increases markedly. Variations in $k$ and $f$ do not change the steady-state behaviour if they remain in the same proportions; however, decreasing both $k$ and $f$ reduces dislocation storage rate and thus delays the initiation of microcracking and limits microcrack growth.

\subsection{Comparison with data}

\begin{figure}
  \centering
  \includegraphics{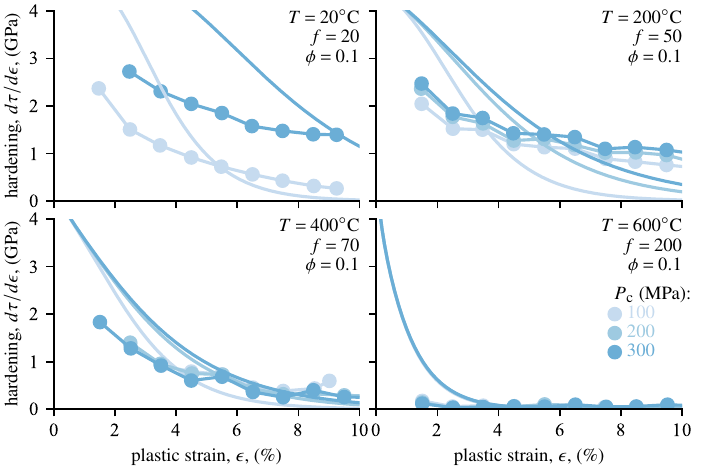}
  \caption{Predicted hardening rates (solid lines) as function of strain in Carrara marble, and comparison with data (circles) from \citet{rybacki21}. Increasing temperature can be simulated by increasing the recovery parameter $f$.}
  \label{fig:hardening_data}
\end{figure}

The model predicts, almost by design, that strain hardening rates decrease with increasing strain, and increase with confining pressure. Hardening rates predicted by the model are of similar orders of magnitude as and follow similar trends to those observed experimentally in Carrara marble by \citet{rybacki21} (Figure \ref{fig:hardening_data}), but are not quantitatively accurate. With the parameter values of Table \ref{tab:param}, and only changing the recovery parameter $f$ at different temperatures, the model tends to overpredict the reduction in hardening rate with increasing strain. In the absence of cracking, the characteristic strain at which hardening drops is of the order of $1/f$, and reduces further if cracking occurs. Decreasing the parameter $f$ does not significantly improve the fit to the data, because the resulting increase in dislocation storage is balanced by a decrease due to additionnal cracking.

\begin{figure}
  \centering
  \includegraphics{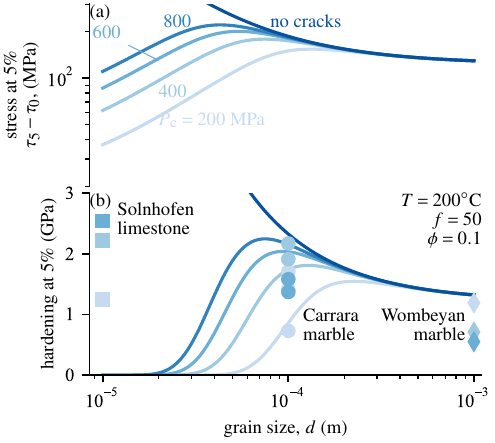}
  \caption{Modelled stress change (a) and hardening rate (b) at 5\% strain as functions of grainsize, and comparison with available experimental data \citep{harbord23}.}
  \label{fig:grainsize}
\end{figure}

Further comparison with experimental data in terms of predicted hardening rates at different grain sizes \citep{harbord23} shows some shortcomings in the modelling approach (Figure \ref{fig:grainsize}). Both strength and hardening rate measured at 5\% strain decrease with increasing grain size at grain sizes above around $100$~$\mu$m, which is broadly consistent with available data in calcite. However, the modelled behaviour at smaller grain sizes shows a decrease in strength and hardening rate that is at odds with experimental data. This discrepancy arises due to the strong dependency of crack growth on pile-up length $L$, which is taken here equal to the grain size: with decreasing grain size, cracks grow more easily, reduce dislocation storage and thus limit hardening rates. It is clear that this mechanism is not correct: experimental results show a clear increase in strength and hardening with decreasing grain size. Unfortunately, experimental observations of crack densities in the semi-brittle regime in fine-grained calcite rocks are scarce, and it is not currently possible to directly probe if cracking is indeed more prevalent in these rocks compared to coarser grained materials. Recent ultrasonic results by \citet{harbord23} suggest that Solnhofen limestone (grain size of the order of $10$~$\mu$m) undergoes more microcracking than Carrara marble (grain size of the order of $100$~$\mu$m) during deformation at high pressure, low temperature, as evidenced by a more marked decrease in wave velocity with increasing strain. Nevertheless, the mechanical data clearly show that fine-grained rocks are stronger and harden at least as much as coarse-grained materials: the effect of cracks is thus more limited than predicted by the model.

A number of possibilities exist to solve the discrepancy, such as: (1) an increase in toughness with decreasing grain size, e.g., due to bridging or cracks stopping at grain boundaries; or (2) a significant decrease in the number of dislocation stored in pile-ups when grain size is small. Both these options would require changing one or several model parameters as a function of grain size, so the model could be modified to produce the required qualitative behaviour. The appropriate modifications would however require further work and independent experimental data, notably in terms of crack density, toughness, and more generally dislocation density.

\section{Discussion and conclusions}

\subsection{Limitations}

The comparison of model predictions with available laboratory data on calcite highlights several important limitations of the modelling approach. Firstly, the quantitative agreement in terms of hardening as a function of strain at different pressures is not great (Figure \ref{fig:hardening_data}). Secondly, the grain size effect is not correctly predicted in fine-grained materials (Figure \ref{fig:grainsize}).

We should keep in mind that the model has been kept as simple and elementary as possible, so that the role of microcracks (initiating and growing from pile-ups, and sinks to mobile dislocations) could be investigated without any further complexity. Even in this simple acception, the model contains many parameters that are not well constrained by independent datasets. This is the case even for the forest hardening parameter $k$ and dynamic recovery parameter $f$, which had to be constrained by deformation and dislocation density data obtained at temperatures higher than those directly relevant to semi-brittle flow. A considerable simplification used here was to consider that a fixed proportion of stored dislocations form pile-ups: this assumption is likely too restrictive. Furthermore, while anecdotal evidence exists for a Bauschinger effect in rocks deformed in the semi-brittle regime, the level of backstress remains poorly known. It was measured recently in olivine at room temperature  and $1250^\circ$C by \citet{hansen19} and \citet{hansen21}, respectively, but it is not known in calcite marble and more systematic data is needed in many rock types.

Despite the wide range of possible parameter values allowed in the absence of firm independent measurements, it is difficult make the model fit all observations, especially the grain size dependence. More complexity is clearly required, with more detailed physics of the microcrack growth mechanisms. Possibilities of extension include toughening mechanisms due to cracks encountering grain boundaries and crack-tip shielding by dislocations \citep[][Chap. 7]{lawn93}. Dislocations ``dragged'' at crack tips have been observed in olivine single crystals \citep{gaboriaud81}. Unfortunately, these processes are hard to quantify in rock forming minerals, and detailed modelling attempts are perhaps premature at this stage.

% what does not work:
% - a lot of options in terms of parameter values
% - cannot reproduce grain size sensitivity without changing other parameters: too much cracking predicted, leading to weakening.
% - twinning: probably relevant at small strain
% - temperature, rate effects: influence on f, but also other recovery mechanisms
% - we have ignored everything else but dislocation motion: friction?
% - cracks may weaken rocks due to other reasons, notably when they interact.

\subsection{Other deformation mechanisms and models}
% connection to HNN
% connection to nicolas's model

We assumed here that all the shear strain was due to dislocation motion. This is an end-member scenario that allowed us to capture the specific feedback of cracking on plastic flow. In nature, more deformation mechanisms co-exist in the semi-brittle regime. Microcracking itself produces additional strain; with purely vertical cracks as assumed here, the additional strain is only in the radial direction and not in the axial direction, and microcracking only contributes to dilational volumetric strain. Twinning contributes to shear strain, and could be included in the model by considering the volume fraction of twins and its evolution with deformation \citep[e.g.][]{bouaziz08}. Experimental data in calcite tend to show that twin spacing is not strongly dependent on strain beyond a few percent axial strain \citep{rybacki13,rowe90}, so only the early stages of deformation should be impacted by this contribution. As discussed earlier and more extensively in \citet{rybacki21}, twinning may lead to additional hardening by reducing dislocation mean-free path: the modelling approach is perfectly suited to include those effects, but more systematic data are needed to fully constrain the appropriate parameter values.

One other mechanism that likely operates in the low pressure regime in parallel to dislocation glide is frictional sliding between neighbouring grains. Frictional slip on preexisting defects, although not directly documented in semi-brittle microstructures, has been extensively studied in the context of micromechanics of brittle deformation in rocks \citep[e.g.][Section 6.2]{paterson05}. It can produce inelastic shear strain, hysteresis in stress-strain behaviour (analogue to backstress effects), and can be the source of tensile microcracks. In the model of \citet{nicolas17}, sliding defects and their associated tensile cracks were included in series with other inelastic processes, and no feedback between frictional slip, cracking and dislocation motion was included. The model proposed here is well suited to include separate cracking processes, arising for instance from frictional slip (so-called ``wing cracks''). This can be done by using a more general internal state variable that represents crack density (instead of our $\ell$ that was completely determined by dislocation pile-ups), and determine an evolution law for this quantity based on all cracking sources. As stated by \citep{wong90}, the mathematics of tensile cracking from dislocation pile-ups is strictly the same as that from sliding frictional defects, because linear elastic fracture mechanics applies regardless of the source of initial slip and stress concentration. It is thus possible to consider other microcracking sources within the same framework.

%By contrast, an earlier model from \citet{horii86} fully included frictional slip, tensile cracking and 

In our model we considered open cracks as dislocation sinks. While it is a plausible hypothesis when we think of dislocations disappearing (and generating steps) at free surfaces, the situation is more complicated if we consider the role of crack tips. In ductile materials, the large stress concentration at elastic crack tips is relaxed by plastic processes, so that cracks tips may be sources of dislocations within grains \citep[as evidenced in olivine and calcite, see for instance][]{gaboriaud81,fredrich89}. It is not clear at this stage how crack tip dislocations can be accounted for in a model: it is possible that those dislocations locally contribute to an increase in toughness, but they may also contribute more generally to stored dislocations.

%The KME approach is phenomenological: it does not rely on a detailed description of dislocation interactions, and the key effects are included in an averaged, scalar way via the dislocation mean-free path and ad-hoc parameters. By contrast, the crack model used here, based on that of \citet{wong90}, treats an idealised crack-dislocation geometry in an exact manner. 

%Recent work by \citet{dong22} has shown that the predictive power of such approaches to explain stress-strain behaviour is quite limited, in part due to the rudimentary treatment of plasticity itself ()

%were based primarily on a description of microscale friction and tensile wing-crack growth: P

\subsection{Outlook}

In an early micromechanical model of the brittle-plastic transition in rocks, \citep{horii86} considered plasticity to be induced by primarily brittle or frictional processes. This was also the case in the approach of \citet{renshaw01}, who used the concept of plastic zones ahead of tensile cracks to determine the nature of fracture growth. In other words, those models considered the effects of plastic flow in a primarily brittle material. Recent work revisiting the approach of \citet{horii86} shows that the contribution of plasticity dominates the overall deformation at large strain \citep{liu23}, which highlights the importance of using a sufficiently realistic plastic flow model. Here, we approached the problem of semi-brittle flow by considering the effect of cracks in a primarily plastic material, where dislocations are the main carrier of deformation as well as the potential source of tensile cracks. 

This approach appears quite promising: The proposed model is capable of reproducing the existence pressure dependency of strength and hardening rate that is commonly observed in the semi-brittle regime. The mechanism suggested here is that dislocation accumulation leads to microcracking, which in turns leads to a weakening effect by allowing mobile dislocations to escape the crystal lattice when they reach the free surfaces of the cracks.

The model is evidently oversimplified and cannot yet reproduce all experimental observations without further parameter adjustments or added complexity. Nevertheless, the KME approach is shown here to be a viable starting point to better understand the semi-brittle regime. Extensions to time-dependent aspects including multiple recovery mechanisms have been shown to be very successful in modelling grain size sensitive plastic flow of olivine \citep{breithaupt21,breithaupt23}. Coupling with twinning processes is also feasible to model calcite marbles \citep{rybacki21,harbord23}.

The source of microcracks was assumed to be dislocation pile-ups, which followed previous work \citep{wong90,nicolas17}. In this approach, the number of crack nuclei per unit volume is equal to the pile-up density, and the only evolving quantity is the average crack length. However, in absence of direct and systematic microstructural observations, this choice is likely overly restrictive. A more general idea could be to include the effect of microcracks using one or more crack density parameters (e.g., crack density tensors as in linear elasticity, \citep{kachanov93}), and make those parameters evolve in a more phenomenological manner, based on a microcrack nucleation and/or growth rate that depends for instance on accumulated plastic strain or other macroscopic quantities. Such an approach might be more flexible, at the cost of fitting ad-hoc phenomenological quantities that could be hard to extrapolate over geological scales. 

In any case, to go beyond basic comparisons of macroscopic stress-strain behaviour, detailed microstructural observations in experimentally deformed samples are required. A rigorous test of the KME approach and its proposed extensions requires systematic, quantitative dislocation density data, together with twin volume, crack density and stress-strain behaviour in rocks deformed in the semi-brittle regime. Perhaps the most critical task is to measure average dislocation density in deformed samples, which is labour-intensive and challenging when dislocations cannot be easily decorated by chemical processes \citep{debresser91}. Modern microscopy techniques such as high-resolution electron backscatter diffraction could be used to determine average densities of geometrically necessary dislocations \citep[e.g.][]{jiang13,wallis19,wallis20}.

%Modern microscopy techniques such as high-resolution electron backscatter diffraction could be 

% the model is far from perfect, and it is evidently too simple to fully account for the complexity of semi-brittle flow in rocks. BUT it captures the key qualitative effect of tensile cracks in the semi-brittle regime.
% The modelling approach is versatile, but we need more systematic data (notably as a function of strain) to go further at this stage.
% implications for nature

\paragraph{Acknowledgments} This work has benefitted from many discussions with Thomas Breithaupt, Brian Evans, Lars Hansen, Chris Harbord, Tongzhang Qu, J\"org Renner, Erik Rybacki, Chris Spiers, and David Wallis. In particular, Lars Hansen provided key comments on an early version of the manuscript. Financial support from the European Research Council under the European Union's Horizon 2020 research and innovation programme (project RockDEaF, grant agreement \#804685 to N.B.), and from the UK Natural Environment Research Council (grant NE/M016471/1) is gratefully acknowledged.

\bibliographystyle{agufull}
\bibliography{references}

\end{document}